\documentclass[reprint,aps,pra,twocolumn,superscriptaddress,showpacs,floatfix,longbibliography,citeautoscript]{revtex4-1}

\usepackage[dvips]{graphicx}

\usepackage{color}

\usepackage{latexsym}

\usepackage{amsmath}

\usepackage{amsthm}

\usepackage{amsfonts}

\usepackage{amssymb}

\usepackage{mathptmx}

\usepackage{subfigure}

\usepackage{mathrsfs}

\usepackage{hyperref}

\hypersetup{colorlinks=true}

\usepackage[all]{hypcap}

\usepackage{gensymb}

\begin{document}
\title{Spontaneous giant exchange bias in antiperovskite structure driven by canted triangular magnetic structure}
\author{Lei Ding}
\email[]{lei.ding.ld@outlook.com}
\affiliation{ISIS Facility, Rutherford Appleton Laboratory, Harwell Oxford, Didcot OX11 0QX, United Kingdom}
\author{Lihua Chu}
\affiliation{State Key Laboratory of Alternate Electrical Power System with Renewable Energy Sources, School of Renewable Energy, North China Electric Power University, Beijing 102206, China.}
\author{Pascal Manuel}
\email[]{pascal.manuel@stfc.ac.uk}
\affiliation{ISIS Facility, Rutherford Appleton Laboratory, Harwell Oxford, Didcot OX11 0QX, United Kingdom}
\author{Fabio Orlandi}
\affiliation{ISIS Facility, Rutherford Appleton Laboratory, Harwell Oxford, Didcot OX11 0QX, United Kingdom}
\author{Meicheng Li}
\author{Yanjiao Guo}
\author{Zhuohai Liu}
\affiliation{State Key Laboratory of Alternate Electrical Power System with Renewable Energy Sources, School of Renewable Energy, North China Electric Power University, Beijing 102206, China.}
\date{04 January 2018}

\begin{abstract}
Exchange bias (EB) refers to a shift of the hysteresis loop along the field axis in materials consisting of ferromagnetic (FM) and antiferromagnetic (AFM) layers, generally after a cooling procedure in high magnetic field. This effect is highly desirable for technological applications ranging from spintronics to magnetic recording. Achieving giant EB effect near room temperature in a small cooling field is thus an on-going technologically relevant challenge for the materials science community. In this work, we present the experimental realization of such a fundamental goal by demonstrating the existence of giant EB after a zero field cooled (ZFC) procedure in antiperovskite Mn$_{3.5}$Co$_{0.5}$N below 256 K. We found that it exhibits an EB field of $-$0.28 T at 50 K after a ZFC procedure accompanied by a large vertical magnetization shift (VMS). Interestingly, this EB field can be elevated up to $-$1.2 T after a cooling procedure with a small applied field of just 500 Oe. Mn$_{3.5}$Co$_{0.5}$N bears the first intermetallic material showing a strong correlation between EB and VMS. We attribute the observed EB effect to a completely new canted triangular magnetic structure determined by neutron diffraction experiment. Finally, we discuss the striking effect of Co substitution on the physical properties of antiperovskite materials and put forward a new strategy for antiperovskite lattice to exploit new single phase materials showing large EB effect at room temperature.
\end{abstract}


\maketitle

\section{Introduction}

\begin{figure*}[t]
\centering
  \includegraphics[width=0.8\linewidth]{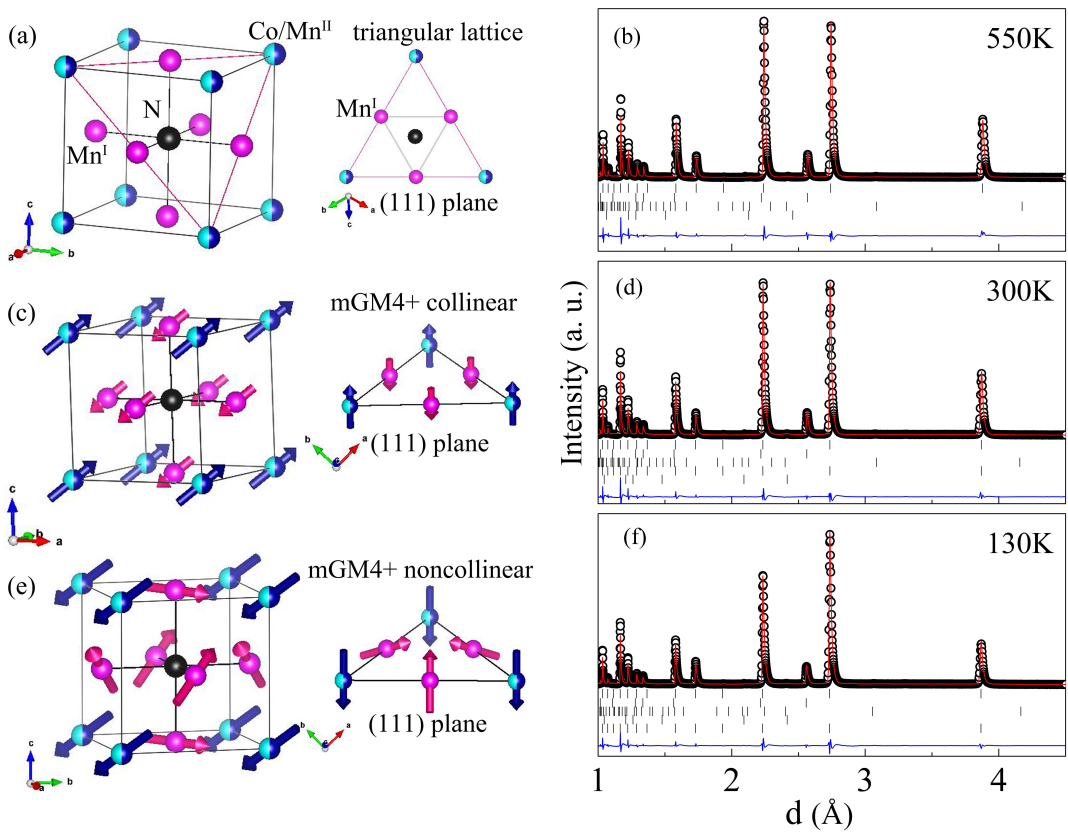}
  \caption{(a), (c) and (e) Nuclear and magnetic structures for Mn$_{3.5}$Co$_{0.5}$N  at 550, 300 and 130 K, respectively. (b), (d) and (f) Observed (open circles) and calculated (line) powder neutron diffraction patterns for Mn$_{3.5}$Co$_{0.5}$N collected at 550, 300 and 130 K. The nuclear reflections are denoted by upper tick marks. The reflections marked in middle belong to impurity phases MnO (3.5(3)\%), Mn$_2$N (0.86(5)\% ) and CoO (0.32(4)\%). The lowest tick marks show magnetic reflections. The bottom line stands for the difference between the observed and calculated patterns.} 
  \label{fgr:1}
\end{figure*}

Exchange bias (EB),\cite{Meiklejohn1957,Nogues2005} which is characterized by a shift of the hysteresis loop along the field axis in materials composed of ferromagnetic (FM) and antiferromagnetic (AFM) layers, possesses significant technological applications in advanced magnetic devices such as spin valves, magnetic tunnel junctions and magnetoresistive sensors.\cite{Skumryev2003,Gibert2012,Morales2009,Gider1998,Nogues1999,Radu2012, Huijben2013,Gilbert2016} It is conventionally generated by a unidirectional exchange anisotropy when the material is cooled in a static magnetic field below the AFM transition.\cite{Gilbert2016,Lage2012,Kuch2006,Miltenyi2000,Nolting2000,Stamps2000,Wang2011} A simultaneous but rarely observed effect is a shift of magnetic hysteresis loop along the magnetization axis, which is referred to as $“$vertical magnetization shift (VMS)$”$.\cite{Nogues2000,Ohldag2003,Giri2011,Yuan2013} Recently, VMS has been observed in several magnetic bilayer systems, in which VMS seems to be an effective source for the exchange bias effect.\cite{Nogues2000,Ohldag2003}

Since the first observation of exchange bias in Co/CoO systems, a large number of investigations have been devoted to the study of EB in nanostructures and heterostructures to develop advanced magnetic materials for practical applications and to microscopically understand it.\cite{Nogues1999,Radu2012, Huijben2013,Gilbert2016,Lage2012,Kuch2006,Miltenyi2000,Nolting2000,Stamps2000,Wang2011}The vast majority of the investigated cases are heterogeneous systems characterized by magnetic phase separations at low temperature. \cite{Giri2011,Nayak2013,Liao2014,Xia2017,Nayak2015}  By contrast, EB effect in single phase bulk material is less exploited.\cite{Giri2011} As a representative system, Heusler Mn-Pt-Ga  has been found to show giant EB higher than 3 T at 5 K after a cooling field of 10 T due to the existence of FM clusters mixed in a ferrimagnetic (FIM) host.\citep{Nayak2015} Nevertheless, such giant effect in Heusler Mn-Pt-Ga occurring at low temperature and requiring a cooling field of 10 T, impedes its practical applications. 
    
Remarkably, the observation of EB effect under zero field cooled (ZFC) procedure, also known as spontaneous exchange bias effect, is less common. Wang $et$ $al.$ first reported such phenomenon in Ni-Mn-In Heusler system,\cite{Wang2011} and later on, it was also disclosed in other intermetallic systems. \cite{Nayak2013,Liao2014,Xia2017,Nayak2015} This effect was explained in terms of a magnetic phase separation between superferromagnetic and AFM regions, in which the strong uniaxial anisotropy needed for EB effect can be obtained isothermally without a field cooled (FC) procedure. 
    
Therefore, the long-standing goal is the quest for giant EB materials, which can operate over a wide temperature range up to room temperature when subject to a relatively small or even null cooling magnetic field. In the present work we suggest a new route to achieve this goal by exploiting the magnetic features of antiperoskite triangular lattice.

Mn-based antiperovskite compounds with formula Mn$_{3}$AX (A=Mn, Ni, Cu, Zn, Ga, Ge, etc.; X=carbon or nitrogen) have attracted great interest as they exhibit a wide variety of novel physical properties, such as negative thermal expansion (NTE),\cite{Takenaka2005,Takenaka2009,Song2011,Wang2012,Ding2011,Tan2015,Lin2015,Chu2018} near zero temperature coefficient of resistivity,\citep{Ding2011APL} spin glass behavior,\citep{Ding2015} barocaloric effect.\citep{Matsunami2015} Most antiperovskite compounds crystallize into $Pm-3m$ space group, as shown in Fig.\ref{fgr:1}, with the magnetic Mn atoms located at the face-centered site, the A atoms at the corner site, and the X atoms at the body-centered site. This arrangement gives rise to a triangular lattice composed of magnetic Mn atoms at the face-centered sites (Fig.\ref{fgr:1}), leading to geometrical frustration. Previous investigations have revealed that all the intriguing properties in Mn-based antiperovskites are attributed to a noncollinear AFM spin configuration on the triangular lattice.\cite{Takenaka2005,Takenaka2009,Song2011,Wang2012} When the corner position of the cubic cell is partially occupied by Mn atoms, novel magnetic properties are expected since one extra magnetic position is involved.\citep{Lin2015,Deng2015} For example, Invar-like behavior has been revealed in Mn$_{3+x}$Ni$_{1-x}$N due to the coexistence of a FM and typical triangular AFM phases.\citep{Deng2015}

\begin{figure}
\centering
  \includegraphics[width=0.6\linewidth]{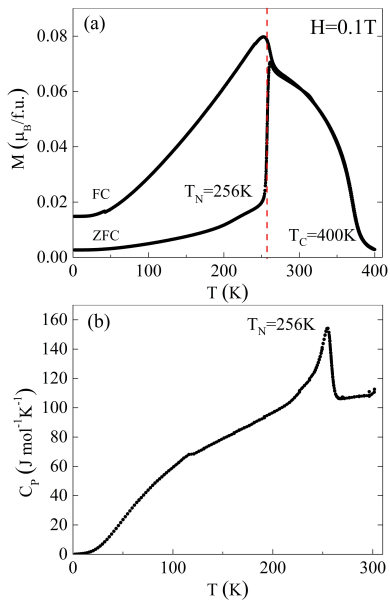}
  \caption{(a) Temperature dependence of the magnetization of Mn$_{3.5}$Co$_{0.5}$N measured with ZFC and FC conditions under 0.1 T from 2 to 400 K. (b) Heat capacity of Mn$_{3.5}$Co$_{0.5}$N in the temperature range of 5-300 K. A very small cusp at 115 K represents the contribution from MnO.} \label{fgr:2}
\end{figure}

In this work, we investigate the structural and magnetic properties of Mn$_{3.5}$Co$_{0.5}$N. For the first time, we observed considerable VMSs below T$_N$=256 K in both FC and ZFC conditions in a magnetic single phase intermetallic system. The VMS, related to the net magnetic moment in nearly compensated FIM, leads to a large EB effect in ZFC condition. Moreover, the value of the EB field can be remarkably increased by applying a small cooling field. The possible mechanism governing this interesting phenomenon is discussed based on the experimentally determined magnetic and nuclear structures. 

\section{Results}

\begin{figure}[t]
\centering
  \includegraphics[width=0.7\linewidth]{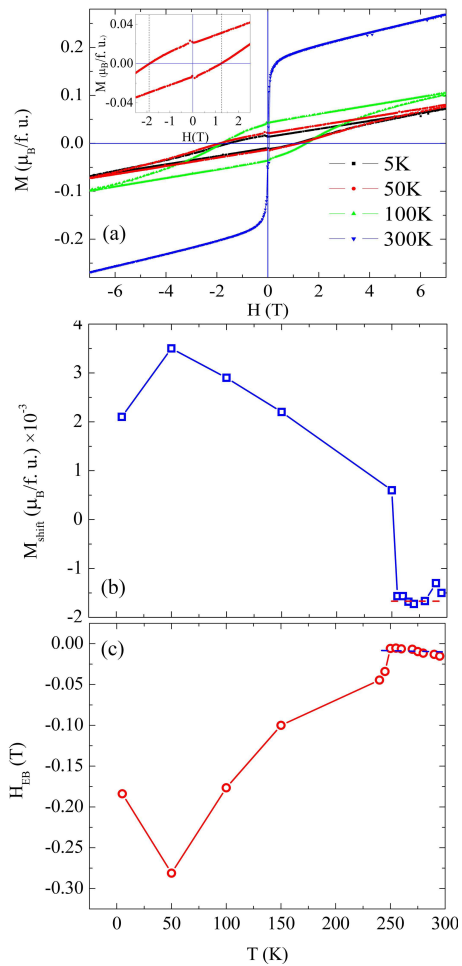}
  \caption{(a) Magnetic hysteresis loops of Mn$_{3.5}$Co$_{0.5}$N at different temperatures under ZFC condition. (b) and (c) Temperature dependence of the H$_{EB}$ and M$_{shift}$ under ZFC condition.} \label{fgr:3}
\end{figure}

The synthesis of the compound has been performed through standard solid state reaction method. Powders of the reactants Mn$_{2}$N and Co were mixed and pressed into pellets. The pellets were then wrapped in Ta foil to prevent oxidation and to capture the N$_{2}$ released during the reaction avoiding the formation of nitride impurities. Synthesis performed with Au foil returned highly oxidized materials as demonstrated in supplementary information. The synthesis conducted with stoichiometric amounts of the reactants, targeting the composition Mn$_{3.5}$Co$_{0.5}$N, results in high concentration of Mn$_{2}$N in the final material, indicating a lack of reagent Co. Syntheses with an excess of Co were then carried out in order to improve the purity of the final material. As summarized in Table S1, in the supplementary information, the best quality samples were achieved when a 110\%wt excess of Co powder has been used. No trace of Co were found in  the final material but there exists a very tiny amount of CoO impurity (0.32(4)\%) in the nanometric range, featured by a non-resolution limited reflection (See Fig. S1). However, EDX and XRD measurements performed on the Ta foil and on the quartz tube surface reveal the formation of CoTa$_{2}$ which explains the high Co losses during the reaction. We present in the main text the results obtained on the sample synthesized with 110\%wt Co excess, whereas magnetization and neutron measurements performed on other samples synthesized with different amounts of Co excess are reported in the supplementary information (see Fig. S7-S9). It is worth noting that the magnetic properties are very similar in all the investigated samples. In particular, Rietveld refinement of the neutron diffraction data, taking advantage of the good contrast between Co and Mn, always returns generally the same nominal composition for the main phase, namely Mn$_{3.5}$Co$_{0.5}$N (Table S2). These observations suggest that the detected physical properties are indeed intrinsic to the Mn$_{3.5}$Co$_{0.5}$N compound.

\begin{figure*}[t]
\centering
  \includegraphics[width=0.8\linewidth]{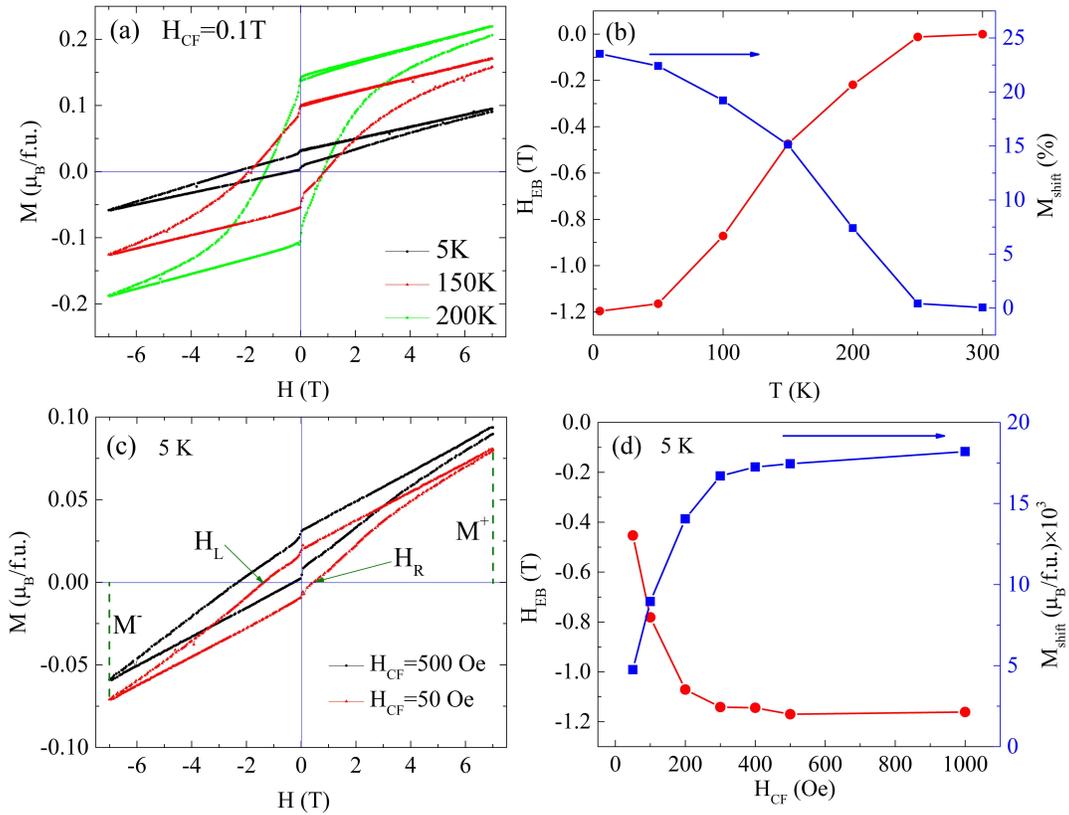}
  \caption{(a) Magnetic hysteresis loops of Mn$_{3.5}$Co$_{0.5}$N at different temperatures under a cooling field of 0.1 T. (b) Temperature dependence of the H$_{EB}$ and M$_{shift}$ under a cooling field of 0.1 T. (c) Magnetic hysteresis loops of Mn$_{3.5}$Co$_{0.5}$N at 5 K under cooling fields of 50 and 500 Oe. (d) Cooling-field-dependent H$_{EB}$ (a) and M$_{shift}$ at 5 K.}
  \label{fgr:4}
\end{figure*}

As shown in Fig. \ref{fgr:1}, neutron powder diffraction indicates that Mn$_{3.5}$Co$_{0.5}$N crystallizes in antiperovskite cubic cell with $Pm-3m$ symmetry, in which Mn and N occupies the 3c and 1b position, respectively, whereas the corner site 1a (0,0,0) is occupied by half Co and half Mn. Therefore the chemical formula could be written as Mn$^{I}_{3}$Mn$^{II}_{0.5}$Co$_{0.5}$N. The corresponding results from Rietveld refinement including the fractional occupation of Mn and Co are presented in Table S3 in the supplementary information. While temperature-dependent neutron diffraction in the temperature range of 8-550 K does not show structural symmetry change, large NTE  was observed from 220 to 260 K, accompanied by a magnetic phase transition at T$_N$= 256 K (Fig. S11). In sharp contras to other antiperovskites where NTE is usually associated with the so-called mGM5+ AFM magnetic structure,\cite{Takenaka2005,Takenaka2009,Song2011,Wang2012} we found  that the NTE in Mn$_{3.5}$Co$_{0.5}$N is in fact related to a more complex magnetic configuration transformed by mGM4+, as explained in the following.

Fig. \ref{fgr:2}(a) shows the temperature dependence of the magnetization of Mn$_{3.5}$Co$_{0.5}$N, measured under 0.1 T in both ZFC and FC conditions. Two magnetic phase transitions are clearly seen: upon cooling, it first undergoes a FIM transition at T$_C$=400 K, then an AFM transition sets in at T$_N$=256 K. The irreversibility between ZFC and FC curves reflects the presence of a weak FM component below T$_N$. Specific heat from 2 to 300 K further supports the presence of an AFM transition at 256 K (Fig. \ref{fgr:2}(b)).  

The magnetic hysteresis loops of Mn$_{3.5}$Co$_{0.5}$N were measured over a temperature range of 5-300 K under ZFC condition, giving particular care to eliminate any small trapped magnetic field in the SQUID at 400 K. The measurements, shown in Fig. \ref{fgr:3} were recorded between ${-}$ 7 T and 7 T using the protocol: 0 $\longrightarrow$ 7 T $\longrightarrow$ 0 $\longrightarrow$ $-$7 T $\longrightarrow$ 0 $\longrightarrow$ 7 T. The hysteresis loops measured below T$_{N}$, see Fig. \ref{fgr:3}(a), show the occurrence of large horizontal and vertical shifts under ZFC condition. Especially, these shifts make the loops asymmetric at 50 K, entailing a giant EB field of H$_{EB}$=${-}$0.28 T. The H$_{EB}$ and vertical magnetization shift M$_{shift}$ are defined as H$_{EB}$=(H$_L$+H$_R$)/2 and M$_{shift}$=(M$^+$+M$^-$)/2, respectively, where H$_L$ and H$_R$ are the left and right coercive fields, and M$^+$ and M$^-$ are the maximum positive and negative magnetization (see the marks in Fig. \ref{fgr:4}(c)). The temperature evolution of H$_{EB}$ and M$_{shift}$ is presented in Fig. \ref{fgr:3}(b-c), showing that large value of M$_{shift}$ always accompanies large negative H$_{EB}$. Note that the large EB occurs in the vicinity of T$_N$=256 K, ruling out the contribution of a very tiny amount impurity phase CoO (AFM below 290 K). Several repeat measurements of the hysteresis loop at 5 K after ZFC always exhibit a small drop of the EB field from 50 to 5 K. This is likely related to the presence of a very tiny amount of CoO as previous works about CoO nanoparticles suggested that there exists magnetic phase transition around 10 K, depending on the size of nanoparticles\cite{Ghosh2005, Dai2013}.

Fig. \ref{fgr:4} shows hysteresis loop measured under field cooled condition, together with the field and temperature dependence of the H$_{EB}$ and of M$_{shift}$. One can immediately see that the robust EB effect occurs only below around 250 K, again excluding the influence of impurity phase CoO. In consistence with the ZFC measurements, the latter two quantities are strictly related: as the VMS changes from 7.5\% at 200 K to 23\% at 50 K, the absolute value of H$_{EB}$ rises typically from 0.22 T to 1 T. The maximum H$_{EB}$ is ${~}$1.2 T at 5 K under a cooling field of only 0.1 T. In order to further study the giant EB effect, we have investigated its cooling field dependence in the magnetic field range 0.005-5 T (results measured under high fields are shown in Fig. S12-S16). It is evident that the value of H$_{EB}$ increases rapidly with increasing the cooling field (H$_{CF}$) and saturates to 1.2 T above H$_{CF}$=500 Oe (see Fig. \ref{fgr:4}(d) and Fig.  S12). Accordingly, M$_{shift}$ rapidly increases from 50 to 500 Oe and reaches its maximum under a cooling field of 0.1 T (see Fig. \ref{fgr:4}(d)). This feature makes Mn$_{3.5}$Co$_{0.5}$N clearly distinct from the documented magnetic phase-separated oxides, super spin glassy or cluster glassy alloys, where H$_{EB}$ typically decays rapidly with high cooling field, known as minor loop effect.\cite{Giri2011}

\begin{figure*}[t]
\centering
  \includegraphics[width=0.8\linewidth]{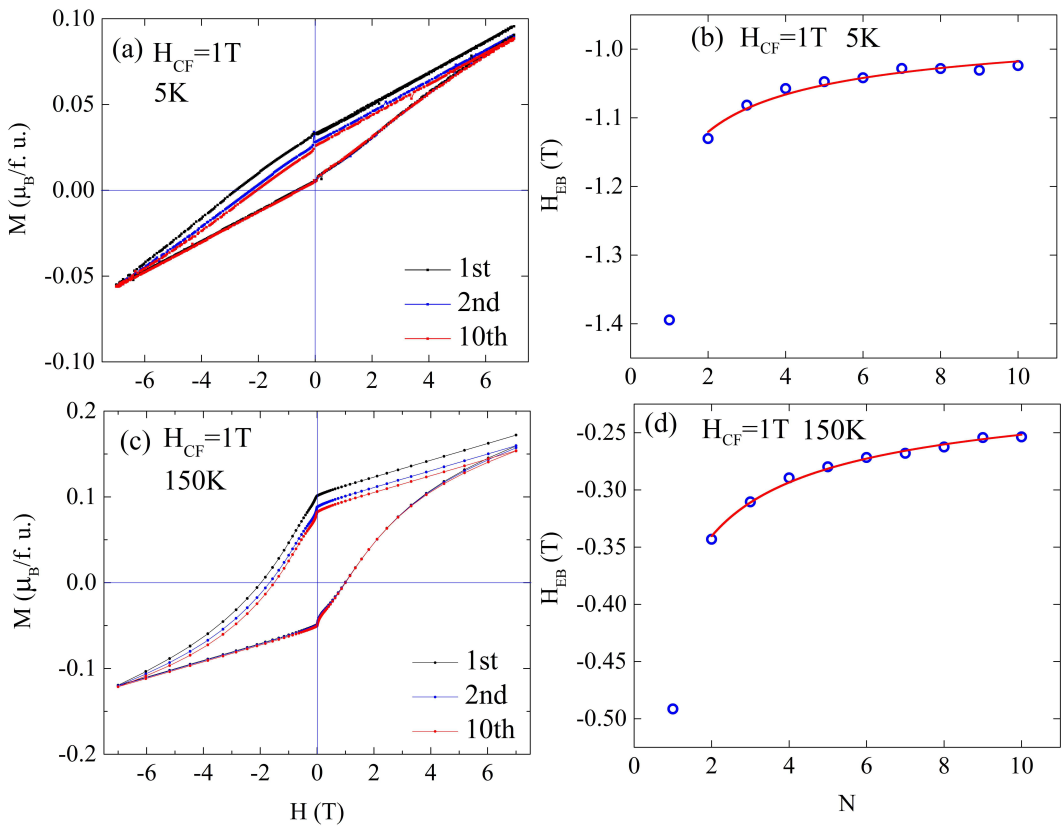}
  \caption{Training effect of Mn$_{3.5}$Co$_{0.5}$N. Consecutive hysteresis loops measured at 5 K (a)and 150 K (c) after field cooling under 1 T. H$_{EB}$ as a function of cycling index number, N, at 5 K (b) and 150 K (d). Solid lines show the best fits with Eq. \ref{eq:1} to the
data for N>1.}
  \label{fgr:5}
\end{figure*}

It is well known that one of the interesting characteristics in exchange bias materials is the so-called training effect, which describes the monotonous decrease of exchange-bias field with the cycling index number N. Fig. \ref{fgr:5} shows the training effect, measured up to N=10, in Mn$_{3.5}$Co$_{0.5}$N at 5 and 150 K after a FC procedure with the applied field of 1 T. At both temperatures, the decrease of H$_{EB}$ between the first and second loops is evident (H$_{EB}$ reduces by ~19\% at 5 K) while from the second to the tenth loop H$_{EB}$ is slightly and gradually reduced. Such a relaxation feature is more clear in Fig. \ref{fgr:5}(b) and (d) which show the  H$_{EB}$ as a function of cycling index number N. The decrease of H$_{EB}$ as a function of cycling index number N (N>1) follows the empirical powder law:

\begin{equation}
 H_{EB}-H^{\infty}_{EB}\propto \dfrac{1}{\sqrt{N}}
 \label{eq:1}
\end{equation}
where H$^{\infty}_{EB}$ is the exchange-bias field in the limit of infinite loops. The powder-law fit yields H$^{\infty}_{EB}$=$-$0.93 T and $-$0.18 T for cycles at 5 and 150 K, respectively.

Our neutron diffraction data are consistent with the magnetic susceptibility and heat capacity results, indicating the onset of long-range magnetic ordering below the relevant critical temperatures T$_C$ and T$_N$. Below 400 K, an enhancement of the nuclear reflections 100 and 110 is observed, whereas neutron diffraction patterns below 260 K indicate an increase of the 110 and 210 reflections. The two sets of magnetic reflections can be indexed by a propagation vector \textbf{k}=0. For both magnetic phases, the best fit was achieved with the model described by the mGM4+ irreducible representation.\cite{Fullprof,Campbell2006,Bilbao} The refined neutron patterns and the corresponding illustrations of magnetic structures are shown in Fig. \ref{fgr:1}. The collinear magnetic structure at 300 K consists of ferromagnetically aligned Mn$^{II}$/Co spins (1.20(2)$\mu_B$) along the easy axis direction [111] and antiparallel Mn$^I$ spins (0.15(5)$\mu_B$), giving rise to a net magnetic moment of 0.75$\mu_B$. In fact, this magnetic configuration is the same as the parent compound Mn$_4$N.\cite{Takei1962} By contrast, the magnetic structure at 130 K is more complex as it is characteristic of canted noncollinear AFM sublattice composed by Mn$^I$ sites, which is partially compensated by ferromagnetically ordered Mn$^{II}$ (Fig. \ref{fgr:1}(e)). The refined moments at 130 K for Mn$^I$ and Mn$^{II}$ are 1.70(6) and 4.4(1) $\mu_B$, respectively.  
 
\section{Discussion}

Let us discuss possible mechanisms underpinning this novel EB in antiperovskite Mn$_{3.5}$Co$_{0.5}$N. The phenomenological model, generally adopted for EB in heterostructures, where the uncompensated AFM spins pin the interfacial FM ones through exchange coupling, is clearly not applicable to Mn$_{3.5}$Co$_{0.5}$N. The EB effect in several alloys and intermetallics has been previously explained by magnetic phase separation or spin glass in materials such as Fe$_2$MnGa, Mn$_2$FeGa and Mn$_{50}$Ni$_{42}$Sn$_8$.\cite{Liu2016,Tang2010,Sharma2015} As mentioned above, a common phenomenon in these systems, which is absent in Mn$_{3.5}$Co$_{0.5}$N, is the minor loop effect, implying a different mechanism with respect to previous studied systems. The FM cluster mechanism in the Heusler Mn-Pt-Ga should also be ruled out since we did not observe any clear change of the irreversibility between ZFC and FC M(T) curves with increasing field (Fig. S17), nor any diffuse scattering contribution in the neutron data indicative of short-range ordering as expected in case of FM clustering or spin glass behavior.\citep{Nayak2015} A similar cooling-field dependence behavior was recently observed in YMn$_{12-x}$Fe$_x$ and YbFe$_2$O$_4$ where global interactions between FM and AFM sublattices are thought to be responsible for their EB effects.\cite{Xia2017,Sun2013} However, they do not display the VMS effect, which is yet significant in producing EB in Mn$_{3.5}$Co$_{0.5}$N, and never observed in an intermetallic system before.

We propose that the giant EB in  Mn$_{3.5}$Co$_{0.5}$N originates from the global interaction between two magnetic sublattices rather than the interfacial exchange coupling. To set up a qualitative model, it is instructive to take into account the magnetic configuration. As presented in Fig.\ref{fgr:1}(e), the magnetic structure of  Mn$_{3.5}$Co$_{0.5}$N at low temperature can be viewed as two magnetic sublattices: one with FM arrangement and the other with canted AFM structure (which yields a FM component antiparallel to the FM sublattice). This configuration is thus analogous to the interface caused by FM/AFM heterostructures but with smaller net magnetic moment. In this structure, the magnetic anisotropy of the AFM sublattice is supposed to be dominant compared to the FM sublattice one and to the exchange coupling between the two sublattices. As a result, when cooled in an external magnetic field, the FM spins rotate according to the external field while the AFM spins remain in the original configuration. However, the canted AFM sublattice generates a non-null magnetic field that imposes on the FM sublattice, leading to the incomplete reversal of FM spins. This small net magnetic moment will act like the pinned spins in the conventional EB model and will cause the VMS as well as the EB effect.\citep{Yuan2013} This may also explain why only a relatively small cooling field is enough to increase substantially the EB. The ZFC EB in  Mn$_{3.5}$Co$_{0.5}$N is also strongly correlated with the VSM, indicating that the strong magnetic anisotropy of AFM sublattice and pinned spins are relevant. 
 
The role of the Co substitution at the corner site of the Mn$_3$AX lattice can be examined by comparing Mn$_{3.5}$Co$_{0.5}$N with other antiperovskite systems where A is occupied by non-magnetic atoms or Mn.\cite{Takenaka2005,Takenaka2009,Song2011,Wang2012} The active contribution of Co is dual: firstly, the presence of magnetic ions at the corner site stabilizes the collinear FIM structure observed below T$_C$ as also reported in the Mn$_4$N parent compound, and secondly, the unique role of Co is to decrease the exchange interactions between the two sublattices and to allow the second AFM transitions associated with mGM4+ noncollinear magnetic structure.  These simple observations put forward new insights into material design of antiperovskites for giant EB around room temperature. A promising way for probing this concept would be the substitution of elements such as Fe and Ni or other non-magnetic elements for Mn$^{II}$ in Mn$_4$N (T$_N$=745 K) in order to decrease the exchange interactions between the two sublattices allowing the EB effect.\citep{Takei1962} Bearing in mind  technological applications, our work also proposes a promising opportunity to realize giant exchange bias in a single thin film of Mn$_{3.5}$Co$_{0.5}$N rather than a conventional complex heterostructure, avoiding the complication and difficulty in the preparation of high quality interfaces.

\section{Conclusions}

To conclude, we have found giant EB effect, up to $-$1.2 T after a FC procedure with a small applied field of 500 Oe, and a large H$_{EB}$=$-$0.28 T in ZFC procedure below T$_N$=256 K in Mn$_{3.5}$Co$_{0.5}$N. Neutron diffraction experiment reveals that Mn$_{3.5}$Co$_{0.5}$N first undergoes FIM transition at T$_C$=400 K, then a complex AFM phase transition with noncollinear spin arrangements occurs at T$_N$=256 K. The latter phase is responsible for the generation of both ZFC and FC exchange bias effects due to the global interaction between two magnetic sublattices, as well as the observed large NTE behaviour. The strong correlation between VMS and exchange bias lays out a distinct approach to yield exchange bias in single-phase materials. These findings provide a new avenue for exploiting advanced magnetic materials and devices.

\section{Experimental Methods}
\subsection{Sample preparation and characterization}
Polycrystalline samples Mn$_{3.5}$Co$_{0.5}$N were synthesized by solid-state reaction between Mn$_2$N and Co powder with an excess of Co. The reaction powders were carefully mixed, ground and pressed into pellets. The pellets were wrapped in Ta foils and sealed in vacuum (P < 10$^{-5}$ $Pa$) into quartz tubes. The excess of cobalt is needed to counterbalance the losses due to the reaction between the Co and the Ta foil leading to the formation of CoTa${_2}$. Then, the pellets were sintered at 800 C for 80 h, and cooled down to room temperature. Magnetic susceptibilities were measured using SQUID-VSM magnetometers (Quantum Design, MPMS3) between 5 and 400 K under magnetic field of 0.1 T with both zero-field-cooled (ZFC) and field-cooled (FC) conditions. Magnetic hysteresis loops were measured at different temperatures using SQUID-VSM from -7 T to 7 T after ZFC and FC conditions. The heat capacity was recorded between 2 and 300 K on cooling at 0 T by a pulse relaxation method using a commercial calorimeter (Quantum Design PPMS). 
\subsection{Neutron diffraction experiments}
Neutron powder diffraction (NPD) experiments were carried out at the ISIS pulsed neutron and muon facility of the Rutherford Appleton Laboratory (UK), on the WISH diffractometer located at the second target station. \citep{Chapon2011} Powder samples (2g) were loaded into 6mm cylindrical vanadium cans and measured in the temperature range of 5-550 K using a Closed Cycle Refrigerator (CCR). Rietveld refinements of the crystal and magnetic structures were performed using the Fullprof program against the data measured in detector banks at average 2$\theta$ values of 58$^{\circ}$, 90$^{\circ}$, 122$^{\circ}$ and 154$^{\circ}$, each covering 32$^{\circ}$ of the scattering plane.\cite{Fullprof} Group theoretical calculations were done using BasIreps, ISODISTORT and Bilbao Crystallographic Server (Magnetic Symmetry and Applications) software.\cite{Campbell2006,Bilbao}

\section{Conflicts of interest}
There are no conflicts to declare.

\section{Acknowledgements}
Lei Ding thanks support from the Rutherford International Fellowship Programme (RIFP). This project has received funding from the European Union' s Horizon 2020 research and innovation programme under the Marie Sk\l{}odowska-Curie grant agreements No.665593 awarded to the Science and Technology Facilities Council. Lihua Chu acknowledges the support from the National Natural Science Foundation of China (NSFC) (No. 11504107). We thank the help received during material characterizations from the Materials characterization laboratory at ISIS facility.

\end{document}